# Epitaxial growth of diluted magnetic semiconductor $Ge_{1-x}Cr_xTe$ with high Cr composition

Y. Fukuma[1], H. Asada[2], S. Senba[3], and T. Koyanagi[2]

[1] *Frontier Research Academy for Young Researchers, Kyushu Institute of Technology, 680-4 Kawazu, Iizuka, Fukuoka 820-8502, Japan*

[2] *Department of Electronic Devices Engineering, Graduate School of Science and Engineering, Yamaguchi University, 2-16-1 Tokiwadai, Ube 755-8611, Japan*

[3] *Ube National College of Technology, 2-14-1 Tokiwadai, Ube 755-8555, Japan*

**Abstract**

IV-VI diluted magnetic semiconductor $Ge_{1-x}Cr_xTe$ layers up to x=0.1 were grown on $SrF_2$ substrates by molecular beam epitaxy. In site reflection high-energy electron diffraction shows a streaky pattern with sixfold symmetry in the plane for the $Ge_{1-x}Cr_xTe$ layer, implying an epitaxial growth of $Ge_{1-x}Cr_xTe$ [111]/$SrF_2$ [111]. A clear hysteresis loop is observed in the anomalous Hall effect measurements due to the strong spin-orbit interaction in the host GeTe. The Curie temperature increases with increasing Cr composition up to 200 K, but there is no clear dependence of the Curie temperature on the hole concentration, implying that the mechanism of the ferromagnetic interaction among Cr ions is different from Mn doped diluted magnetic semiconductors.

Ferromagnetic order in diluted magnetic semiconductors (DMSs) which are compounded semiconductor doped with magnetic ions has provided a wealth of scientific interests and potential technological applications.[1] The much work has been conducted on III-V DMSs because of relatively high Curie temperature ($T_C$) of around 200 K in DMSs.[2] $T_C$ increases with increasing magnetic ions in the host semiconductor, however, a low solubility limit of magnetic ions into III-V semiconductors hampers realizing high-$T_C$. Recently, theoretical works based on a density functional theory suggest that IV-VI DMSs such as $Ge_{1-x}Mn_xTe$ and $Ge_{1-x}Cr_xTe$ are a good candidate for realizing the room temperature ferromagnetism because of its high solubility of magnetic ions in the host semiconductor.[3-5] In addition, GeTe is known as a phase change material between the amorphous phase and the crystalline phase by heating, which is utilized for developing a non-volatile memory.[6] The phase change properties are also observed for Mn-doped GeTe systems and then the magnetic order is also changed from paramagnet in the amorphous state to ferromagnet in the crystalline state.[7-9] More recently, the nonmagnetic GeTe is known as a topological insulator,[10] and giant magnetoresistance is observed for the phase-change $GeTe/Sb_2Te_3$ multilayered structure.[11] Therefore, the magnetic and nonmagnetic heterostructures with strong spin-orbit coupling such as $Ge_{1-x}Mn_xTe/GeTe$ and $Ge_{1-x}Cr_xTe/GeTe$ may offer unique opportunities for studying new spin transport and spin-to-charge current conversion phenomena at the interface.

So far we have grown $Ge_{1-x}Mn_xTe$ and $Ge_{1-x}Cr_xTe$ epilayers on $BaF_2$ substrates by using molecular beam epitaxy.[12,13] Relative high $T_C$ of 190 K and 180 K is realized for $Ge_{0.92}Mn_{0.08}Te$ and $Ge_{0.94}Cr_{0.06}Te$, respectively. According to the theoretical prediction,[3-5] a strong ferromagnetic interaction via the double exchange mechanism is expected for $Ge_{1-x}Cr_xTe$, therefore, higher $T_C$ is expected for further increasing Cr composition. In this study, we grew $Ge_{1-x}Cr_xTe$ epilayers on $SrF_2$ substrates due to the smaller lattice mismatch of the $Ge_{1-x}Cr_xTe/SrF_2$ system compared with the $Ge_{1-x}Cr_xTe/BaF_2$ system, especially for a high Cr composition regime because the lattice constant of $Ge_{1-x}Cr_xTe$ decreases with decreasing Cr composition.[14] A bulk GeTe has rhombohedral structure with the lattice constant of 0.598 nm and the distorted angle of 88°.[15] $SrF_2$ and $BaF_2$ have calcium fluoride structure with the lattice constant of 0.580 nm and 0.620 nm, respectively. In spite of the different crystal structure, the alkaline-earth-fluoride substrates are suited for an epitaxial growth of IV-VI semiconductors.[16] First of all, in order to check crystalline qualities of GeTe on the $SrF_2$ substrate, 200 nm-thick films were grown at various substrate temperatures by using a GeTe compound effusion cell. The cell temperature was kept at 450 °C and the evaporation rate was around 3.3 nm/min. In site reflection high-energy electron diffraction (RHEED) showed a streaky pattern with a characteristic feature of rocksalt structure and a sixfold symmetry in the plane, implying a perfect epitaxial growth of GeTe [111]/$SrF_2$ or $BaF_2$ [111]. Figure 1 shows x-ray diffraction (XRD) patterns of the GeTe layers grown at various substrate temperatures $T_S$. Beside the strong 222 peak of the $BaF_2$ substrate a clear 222 peak of the GeTe layer is visible. Despite of lager lattice mismatch for the GeTe layer on the $SrF_2$ substrate, no significant change of the peak position of the GeTe layer is observed. The lattice constant deduced from the XRD peak position is 0.615 nm for $T_S$ = 300 °C, 350 °C and 0.605 nm for $T_S$ = 250 °C, respectively.

For the growth of $Ge_{1-x}Cr_xTe$ layers with high Cr composition which is far above thermal equilibrium, lower temperature growth was used and additional Cr and $Te_2$ beam fluxes were supplied by an electron beam evaporator. The Te/Cr flux ratio was fixed at around 1.0. Figure 2 shows RHEED patterns along the [$\bar{1}$10] azimuth for the $Ge_{1-x}Cr_xTe$ layer with x=0.05, 0.10 and 0.16 grown on the $SrF_2$ substrate at $T_S$ = 250 °C. A streaky RHEED pattern gradually changed to a spotty RHEED pattern with increasing Cr composition, implying that surface morphology is getting rough. However, for x=0.05 and 0.10, there is no indication of the formation of second phases. For x=0.16, a combination of halo and ring patterns shows a polycrystalline film.

To measure the electrical properties of the $Ge_{1-x}Cr_xTe$ layer, a Hall-bar geometry with 1 mm length and 200 μm width was fabricated by using conventional photolithography and chemical wet etching. The current is applied along the (110) direction. In ferromagnetic materials, the Hall resistivity $\rho_{Hall}$ is expressed as

$$\rho_{\text{Hall}} = R_0 H_Z + R_S M_Z,^{17} \tag{1}$$

where $R_0$ is the ordinary Hall coefficient, $H_Z$ is the magnetic field applied perpendicular to the film plane, $R_S$ is the anomalous Hall coefficient, $M_Z$ is the magnetization. The anomalous Hall term which is the second term of the Eq. (1) is caused by the spin-orbit interaction and allows us to detect magnetic properties of the $Ge_{1-x}Cr_xTe$ layer in the electrical measurements.[12] Figure 3 shows the Hall resistivity as a function of the applied field at low temperatures for x=0.05, 0.10 and 0.16 grown at $T_S$ = 250 °C. Square hysteresis loops are observed for x=0.05 and 0.10, implying that the $Ge_{1-x}Cr_xTe$ epilayers on the $SrF_2$ substrate show a perpendicular magnetic anisotropy. The spontaneous magnetization decreases with increasing temperature and the Curie temperature estimated is around 150 K and 200 K for x= 0.05 and x= 0.10, respectively. However, the polycrystalline $Ge_{1-x}Cr_xTe$ layers with x = 0.16 does not show a ferromagnetic hysteresis loop. In DMSs, a strong interplay between magnetic and transport properties is expected due to (s)p-d exchange coupling. The host semiconductor of GeTe has a narrow bandgap of 0.2 eV with a strong spin-orbit interaction, and therefore a clear ferromagnetic order in $Ge_{1-x}Cr_xTe$ and $Ge_{1-x}Mn_xTe$ can be detected in the anormalous Hall effect. The lack of the magnetic order for x = 0.16 in the anomalous Hall measurement suggests that the second phases such as ferromagnetic $Cr_3Te_4$ with hexagonal NiAs structure or an inhomogeneity of the magnetic impurities could be attributed to the magnetic behavior.

One of characteristic feature in DMSs is a hole-mediated ferromagnetism, that is, the manipulation of magnetic properties such as the Curie temperature, saturation magnetization and magnetic anisotropy by changing the hole concentration.[1] For $Ge_{1-x}Mn_xTe$, the Curie temperature was clearly controlled by the hole concentration because the Ruderman-Kittel-Kasuya-Yosida interaction gives rise to the ferromagnetic coupling between Mn ions.[18-22] The magnetic ground state is antiferromagnetic due to the super exchange interaction. On the other hand, a ferromagnetic ground state with strong exchange interaction in $Ge_{1-x}Cr_xTe$ is theoretically expected because of the double exchange interaction.[3-5, 23] In order to check the hole concentration dependence of the magnetic properties in $Ge_{1-x}Cr_xTe$ epilayers, Ge vacancies which may act as a doubly charged acceptor were changed by the growth condition. Figure 4 shows the Cr composition dependence of the Curie temperature and the hole concentration for the $Ge_{1-x}Cr_xTe$ layers grown at various substrate temperatures. No significant change of the hole concentration on the substrate temperature is observed. In the case of $Ge_{1-x}Mn_xTe$, since the 3d states of Mn are located in energetically lower than the 5p states of Te the hole and magnetic concentrations are independently controlled by the Ge vacancies.[13, 20] For $Ge_{1-x}Cr_xTe$, the $e_g$ orbital in the 3d states of Cr is located at the Fermi level,[5] and then Cr ions substituting in the Ge site in the host GeTe act as an acceptor. The hole concentration may be determined by the competition between the Ge vacancies and the Cr ions. An important finding here is that there is no clear dependence of the Curie temperature on the hole concentration: at around 0.05, the Curie temperature and the hole concentration are 4 K and $2.7\times10^{21}$ cm$^{-3}$, 175 K and $1.3\times10^{21}$ cm$^{-3}$, 110 K and $2.9\times10^{21}$ cm$^{-3}$, for $T_S$ = 300 ˚C, 250 ˚C, 200 ˚C, respectively. Therefore, the mechanism of the ferromagnetic interaction is different between Ge$_{1-x}$Cr$_x$Te and Ge$_{1-x}$Mn$_x$Te, which is consistent with the first principle calculation. The Curie temperature increases with increasing Cr composition up to 200 K, which is higher than that of the Ge$_{1-x}$Mn$_x$Te epilayer. For further increasing the Cr composition as well as the Curie temperature, precise control of the Te/Cr flux ratio during the deposition could be a key parameter because the formation energy of Cr ions substituting in Ge ions in the host GeTe is lower under the Te-rich growth condition.[5]

In summary, we have grown Ge$_{1-x}$Cr$_x$Te epilayers on SrF$_2$ substrates by using molecular beam epitaxy. The Cr composition increases up to x=0.1, compared with that of x=0.06 on BaF$_2$ substrates. The Curie temperature increases with increasing Cr composition and the relative high value of 200 K in DMSs is realized. There is no clear dependence of the Curie temperature on the hole concentration, implying that the mechanism of the ferromagnetic interaction among magnetic ions is different between Cr- and Mn-doped DMSs. Even through a low hole concentration, the interaction among Cr ions into GeTe may be ferromagnetic. Such a unique features in Ge$_{1-x}$Cr$_x$Te epilayers would offer an in interesting opportunity to study magnetotransport properties with breaking time-reversal symmetry in IV-VI heterostrutures.

**Acknowledgement**

This work was partly supported by ube industries ltd. foundation and Grant-in-Aid for Scientific Research from the Ministry of Education, Culture, Sports, Science and Technology, Japan.

**Figure captions**

**Fig. 1.** X-ray diffraction patterns of (222) reflection for GeTe layers on BaF$_2$ or SrF$_2$ substrate. The substrate temperature is changed from 250 ˚C to 350 ˚C.

**Fig. 2.** RHEED patterns along the [110] azimuth of Ge$_{1-x}$Cr$_x$Te layers with (a) x=0.05, (b) x=0.10 and (c) x=0.16.

**Fig. 3.** Hall resistivity as a function of the magnetic field applied perpendicular to the film plane for Ge$_{1-x}$Cr$_x$Te layers with (a) x=0.05, (b) x=0.10 and (c) x=0.16.

**Fig. 4.** Cr composition dependence of Curie temperature and hole concentration for Ge$_{1-x}$Cr$_x$Te epilayers.

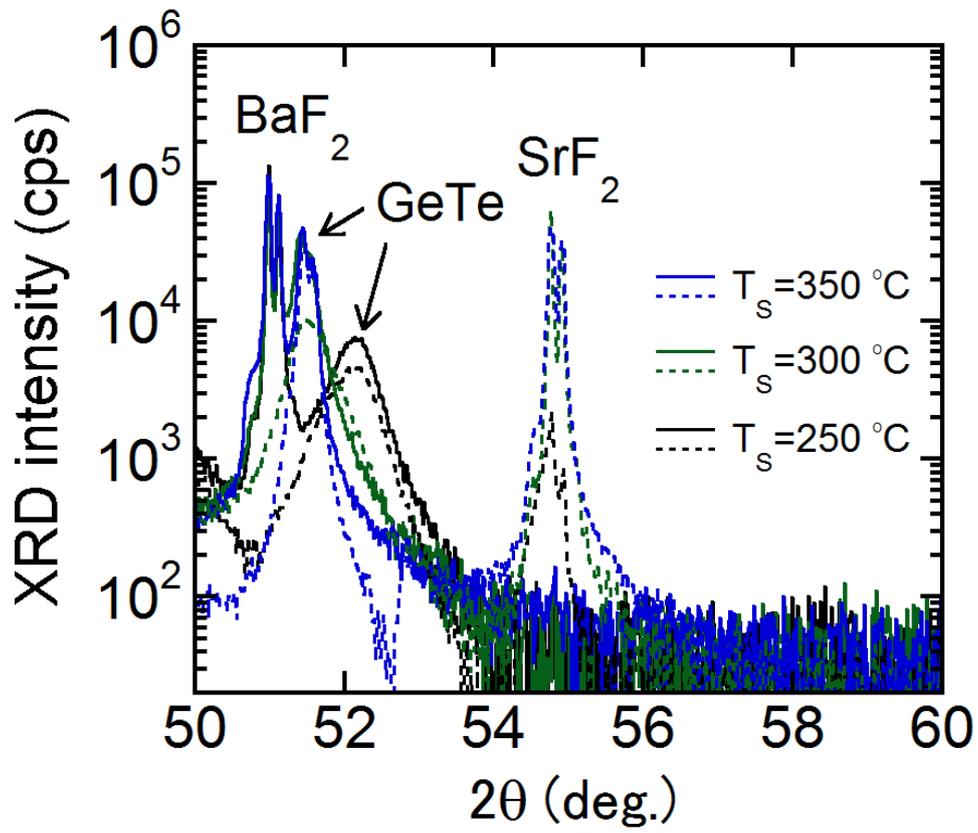

Fig.1 Fukuma *et al*.

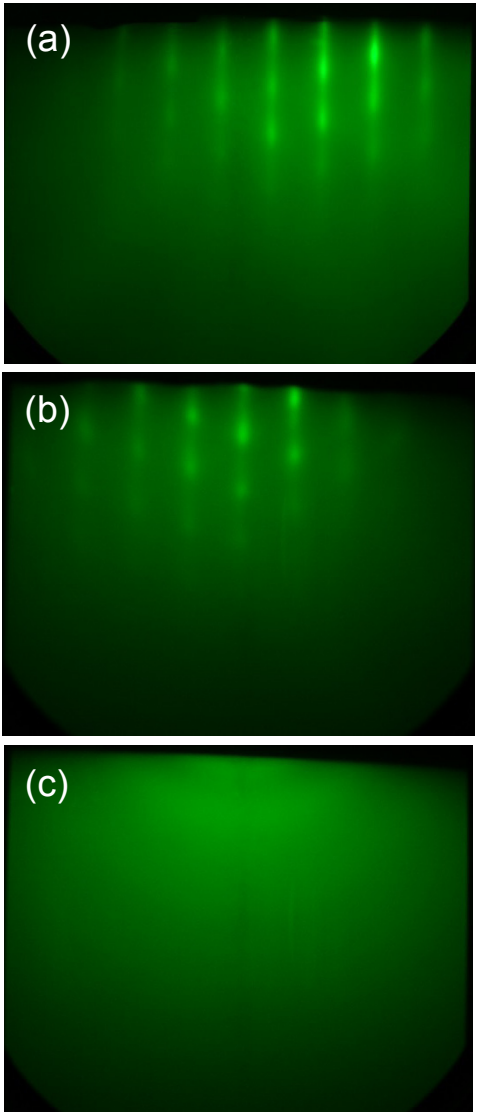

Fig.2 Fukuma *et al*.

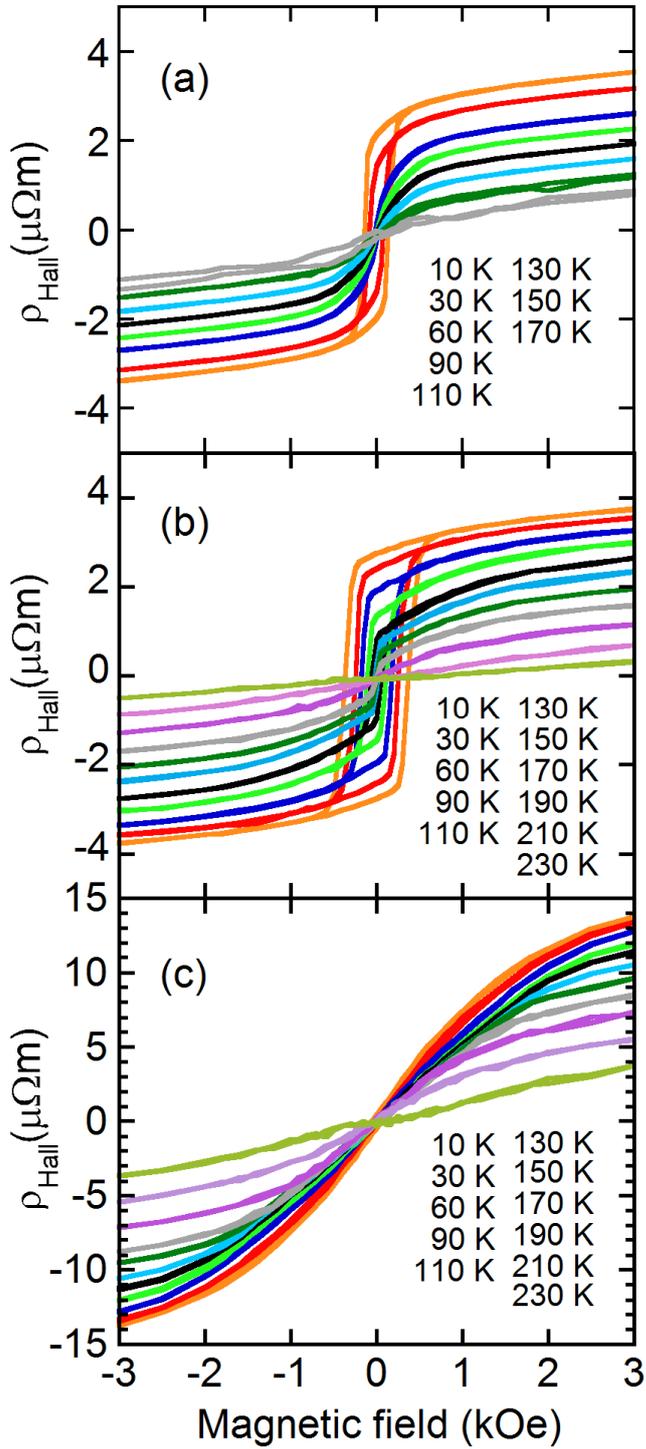

Fig.3 Fukuma *et al*.

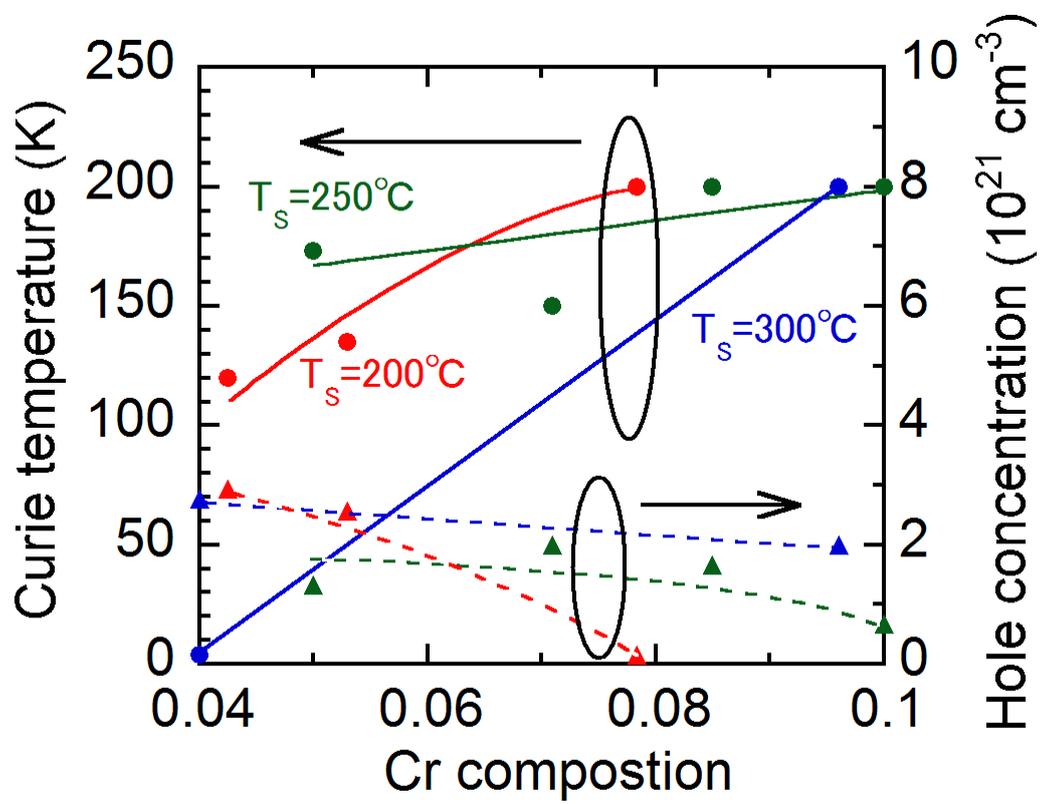

Fig.4 Fukuma *et al*.